\documentclass[dvipdfmx,twocolumn, superscriptaddress]{revtex4}
\usepackage{graphicx}
\usepackage{xcolor}
\usepackage{pdfpages}
\usepackage{amssymb}
\usepackage{amsmath}
\usepackage{amsbsy}
\usepackage{color}
\usepackage{ulem}
\usepackage{epstopdf}

\def\vec#1{\mbox{\boldmath $#1$}}

\begin{document}
\title{Quantifying the mobility of chromatin during embryogenesis: Nuclear size matters}

\author{Aiya K. Yesbolatova}
\affiliation{Department of Genetics, School of Life Science, The Graduate University for Advanced Studies, SOKENDAI, Mishima 411-8540, Japan}
\affiliation{Cell Architecture Laboratory, Department of Chromosome Science, National Institute of Genetics, Mishima 411-8540, Japan}
\author{Ritsuko Arai}
\affiliation{Cell Architecture Laboratory, Department of Chromosome Science, National Institute of Genetics, Mishima 411-8540, Japan}
\affiliation{Present address: Department of Anatomy and Histology, Fukushima Medical University, School of Medicine, 1 Hikarigaoka, Fukushima, 960-1295, Japan}
\author{Takahiro Sakaue}
\email{sakaue@phys.aoyamna.ac.jp}
\affiliation{Department of Physics and Mathematics, Aoyama Gakuin University, 5-10-1 Fuchinobe, Chuo-ku, Sagamihara, Kanagawa 252-5258, Japan}
\author{Akatsuki Kimura}
\email{akkimura@nig.ac.jp}
\affiliation{Department of Genetics, School of Life Science, The Graduate University for Advanced Studies, SOKENDAI, Mishima 411-8540, Japan}
\affiliation{Cell Architecture Laboratory, Department of Chromosome Science, National Institute of Genetics, Mishima 411-8540, Japan}


\maketitle

{\bf Chromatin moves dynamically inside the cell nucleus, and its motion is often correlated with gene functions such as DNA recombination and transcription. A recent study has shown that during early embryogenesis of the nematode, {\it Caenorhabiditis elegans}, the chromatin motion markedly decreases. However, the underlying mechanism for this transition has yet to be elucidated. We systematically investigated the impact of nuclear size to demonstrate that it is indeed a decisive factor in chromatin mobility. To this end, we established a method to quantify chromatin motion inside the nucleus, while excluding the contribution of the movement of the nucleus itself, which allowed us to extract the intrinsic mean-squared displacement (iMSD) of individual chromosomal loci in moving nuclei from the correlated motion of two loci. 
We show that a simple theoretical description, which takes into account the topological constraints of chromatin polymers, can quantitatively describe the relationship between the nucleus size and the chromatin motion {\it in vivo}. Our results emphasize a regulatory role of nuclear size in restricting chromatin motion, and a generic polymer physics model plays a guiding role in capturing this essential feature.}

How does chromatin move? This question has attracted broad interest in both the fundamental and applied sciences, from biology to physics. To address this question, a standard method is a live-cell tracking, in which the real-time motion of fluorescently labeled individual chromatin loci under a microscope are tracked~\cite{Robinett_JCB1996}.  In analyzing the stochastic trajectory obtained ${\vec r}(t)$, a common practice is to calculate the mean-square displacement (MSD): $ \langle ({\vec r}(t_0+\tau) - {\vec r}(t_0))^2\rangle$, where the averaging occurs over the time of origin $t_0$ and/or the different ensemble~\cite{Barkai_PT2012}. 
Previous experiments have reported that, in many cases,  the MSD of chromatin loci exhibits a power-law dependence on the lag-time $\tau$;
\begin{eqnarray}
{\rm MSD}(\tau) = A \tau^{\alpha} , 
\label{eq:MSD}
\end{eqnarray}
where the MSD exponent typically falls in the range $0.3 \lesssim \alpha  \lesssim 0.5$~\cite{Bronstein_PRL2009, Weber_PRL2010, Weber_PNAS2012, Bancaud_GR2013}. An immediate question follows:  why is $\alpha$ so small? The fact that $\alpha <1$ indicates that the diffusion of chromatin loci is anomalous, and there have been intense efforts to clarify the physical mechanism responsible for this observation. Several models have been proposed to explain the observed MSD exponent $\alpha$ taking into account the role of, for example, the viscoelasticity inside the nucleus, the metabolic activity in living cells, or the topological constraints (TCs) relevant for long polymers~\cite{Amitai_PRE2013, Vandebroek_PRE2015, Sakaue_SoftMatter2017, Tamm_PRL2018, Put_PRE2019, Rubinstein_Macro2016}.

Compared to the MSD exponent $\alpha$, less attention has been paid to the amplitude or {\it mobility} $A$ in Eq.~(\ref{eq:MSD}). An earlier observation of telomeres in human chromosomes reported a large variability in $A$ from one telomere to another~\cite{Bronstein_PRL2009}, which may be attributed to environmental heterogeneity. A more recent experiment has reported that the dynamics of the loci become slower as the cell division proceeded during embryogenesis of {\it C. elegans}~\cite{Arai_SR2017}, which may be quantified through $A$.
These observations call for a thorough investigation into the mechanisms controlling the mobility $A$ and its biological implications. The significant reduction in mobility observed during embryogenesis may be related to the global nuclear organization involved in differentiation.

In the present study, we aimed to investigate the relationship between chromatin mobility and the size of the nucleus {\it in vivo}. We adopted the {\it C. elegans} embryo as a model experimental system, in which the nuclear size decreases naturally through cell division, or can be changed artificially by genetic manipulation~\cite{Hara_MBoC2013}. A major difficulty of tracking experiments using embryonic cells lies in the fact that the nucleus itself moves actively, which hampers the quantification of the intrinsic chromatin mobility relative to the container (the nucleus). 
To overcome this problem, we propose a method {\it two-point correlation tracking}, with which one can extract various dynamical information, including the intrinsic chromatin mobility. We elucidate the physical mechanism behind the nuclear size-dependent chromatin mobility by providing a simple theoretical description based on the notion of TC, the predictions of which is in good agreement with experimental results.

\begin{figure}[bh]
\includegraphics[width=0.73\linewidth]{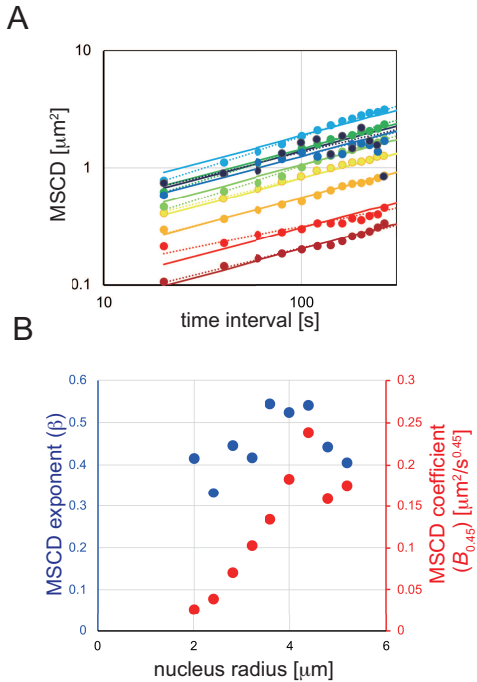}
\caption{Chromatin mobility correlates with nuclear size. (A) MSCD vs. time interval. Different colors represent different sizes of nuclei. Nuclear radius $=$ 2.0 (brown), 2.4 (red), 2.8 (orange), 3.2 (yellow), 3.6 (light green), 4.0 (green), 4.4 (light blue), 4.8 (blue), and 5.2 (navy blue) [$\mu$m]. The solid lines represent fitting to Eq. (3), whereas the dotted lines are fitting to Eq. (3) with $\beta$ fixed to 0.45. (B) The MSCD exponent ($\beta$) is shown in blue. There was no significant correlation between $\beta$ and the nuclear size ($p = 0.3$, regression analysis). The average value of $\beta$ was 0.45. The MSCD coefficient ($B_{0.45}$) is shown in red. There is a statistically significant correlation between $B_{0.45}$ and the nuclear size ($p = 0.002$, regression analysis).}
\label{Fig1}
\end{figure}

{\it Relationship between chromatin mobility and nuclear size.---}
In this study, we imaged the movement of the {\it lacO} locus integrated into a pair of sister chromosomes in the {\it C. elegans} genome using the LacI-GFP fusion protein, as reported previously ~\cite{Bilgir_G32013, Arai_SR2017}. As already mentioned, the tracking analysis of a single locus may lead to an erroneous interpretation, because of a possible influence caused by the translational and rotational motion of the nucleus itself. To exclude such an extrinsic effect, we tracked the positions ${\vec r}_1(t)$ and ${\vec r}_2(t)$ of the pair of {\it lacO} spots in sister chromosomes, and calculated at each time step the distance $d(t) = \sqrt{({\vec r}_2(t) - {\vec r}_1(t))^2}$ between them. From the time series of $d(t)$, we deduced the mean square change in distance (MSCD) as a function of lag-time $\tau$;
\begin{eqnarray}
{\rm MSCD}(\tau; d_0) =  \langle (d(t_0+\tau) - d(t_0))^2\rangle . 
\label{MSCD}
\end{eqnarray}
where $d_0 = d(t_0)$ is the initial distance~\cite{MineHattab_NCB2012, Arai_SR2017}. Fig. 1 summarizes the results for wild-type cells, where we take the average over $d_0$. Similar to the MSD (Eq.~(\ref{eq:MSD})), the MSCD can be fitted with a power-law scaling
\begin{eqnarray}
{\rm MSCD}(\tau) = B \tau^{\beta}.
\label{eq:MSCD}
\end{eqnarray}
Previously, we reported that the MSCD exponent $\beta$ is almost insensitive to the early embryonic stage from 2-cells to 48-cells, but the amplitude $B$ decreases as the stage progresses~\cite{Arai_SR2017}. Because embryonic development is accompanied by a reduction in nuclear size, we hypothesized that the reduction in $B$ was induced by the reduction in nuclear size. In the present study, to focus on the relationship between the nuclear size (radius $R$) and chromatin mobility (MSCD), we first grouped nuclei of similar sizes (i.e., $0.4 \times i - 0.2 \le R \  [\rm{\mu m}] < 0.4 \times$ $i + 0.2$, $i = 1,2,3...$) (Fig. 1A). We fitted the MSCD data from each group of nuclei to Eq.~(\ref{eq:MSCD}). The MSCD exponent $\beta$ was almost insensitive to $R$ (Fig. 1B (blue), $p$ = 0.3), consistent with our previous report~\cite{Arai_SR2017}. We fitted the MSCD data for each group of nuclei to Eq.~(\ref{eq:MSCD}) by fixing the exponent to the average value $\beta = 0.45$. The amplitude obtained ($B$) exhibited a clear trend against nuclear size, in that it takes a smaller value for smaller $R$ (Fig. 1B (red), $p$ = 0.002). These results demonstrate a correlation between chromatin mobility and nuclear size, a finding which is consistent with the previously established correlation between chromatin mobility and the early embryonic stage~\cite{Arai_SR2017}.

{\it Isolating the impact of nuclear size---}
To verify whether changes in nuclear size can cause changes in chromatin motion in a cell-stage independent manner, we induced changes in nuclear size using genetic manipulation.  RNA-mediated interference (RNAi) of {\it ima-3} and {\it C27D9.1} genes induce smaller and larger nuclear sizes, respectively ~\cite{Hara_MBoC2013, Weber_CB2015}. We focused on the 8-cell stage, which may correspond to a transition from high to low mobility of the chromatin~\cite{Arai_SR2017}. The RNAi of the {\it ima-3} and {\it C27D9.1} genes widened the range of nuclear size (Fig. 2A). When we plot the amplitude of the chromatin motion $B$ with the fixed exponent of $\beta = 0.45$ in Eq.~(\ref{eq:MSCD}) as shown in Fig. 1B, but only from the eight-cell stage nuclei, we detected a clear correlation between the chromatin motion and the nuclear size (Fig. 2B, Sup. Fig. S1). These results strongly support our hypothesis that chromatin mobility depends on nuclear size.

\begin{figure}[bh]
\includegraphics[width=0.73\linewidth]{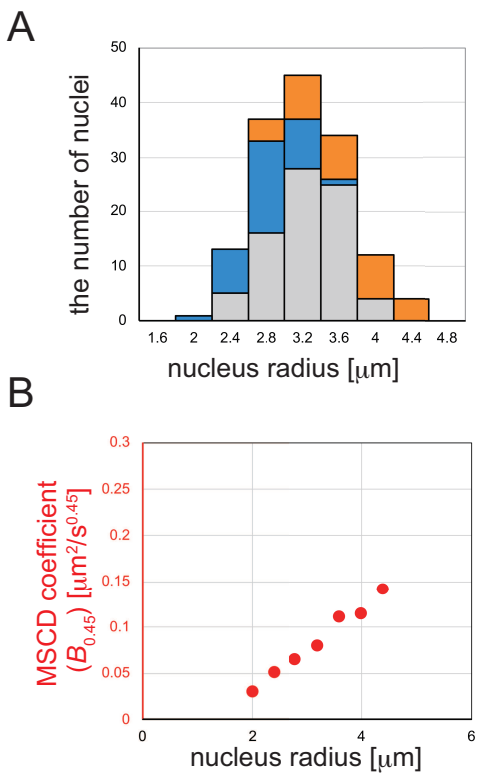}
\caption{Changes in nuclear size can induce changes in chromatin motion. (A) Distribution of the radii of nuclei in the 8-cell stage. Control (gray), \textit{ima-3} (RNAi) (blue), and \textit{C27D9.1} (RNAi) (orange). (B) Correlation between the MSCD coefficient and the nuclear radius in the 8-cell stage ($p = 10^{-5}$, regression analysis).}
\label{Fig2}
\end{figure}

{\it Two-point correlation tracking.---}
Our two-point tracking experiments allowed us not only to calculate the MSCD, but also to evaluate the contribution of the motion of the nucleus in the "observed MSD (oMSD)" from the dynamic correlation of two loci.
Let ${\vec d}(t) = {\vec r}_2(t) - {\vec r}_1(t)$ be a vector connecting two loci points at time $t$. Its increment $\Delta {\vec d}(\tau) = {\vec d}(t+\tau) - {\vec d}(t)$ during a time interval $\tau$ is expressed as the sum of the displacements of two loci during the same interval; that is, $\Delta {\vec d}(\tau)  = \Delta {\vec r}_2(\tau) - \Delta {\vec r}_1(\tau)$, where $\Delta {\vec r}_i (\tau) = {\vec r}_i(t+\tau) - {\vec r}_i(t)$ for $i = 1, 2$. Thus, the mean squared change in this vector can be decomposed as $\langle (\Delta {\vec d}(\tau))^2 \rangle = \langle \Delta {\vec r}_1(\tau)\rangle^2 + \langle \Delta {\vec r}_2(\tau)\rangle^2 - 2 \langle \Delta {\vec r}_1(\tau) \cdot  \Delta {\vec r}_2(\tau) \rangle$, which can be rearranged as
\begin{eqnarray}
 {\rm oMSD}(\tau) = \frac{\langle (\Delta {\vec d}(\tau))^2 \rangle}{2}  +  \langle \Delta {\vec r}_1(\tau) \cdot  \Delta {\vec r}_2(\tau) \rangle, 
\label{MSCDV}
\end{eqnarray}
If the nucleus itself is quiescent, the displacement correlation $\langle \Delta {\vec r}_1(\tau) \cdot  \Delta {\vec r}_2(\tau) \rangle$ between two loci will become negligibly small compared to oMSD~\cite{Zidovska_PNAS2013,Michieletto_MacrLett2021}. In this case, the mean square change in ${\vec d}(t)$ is reduced to approximately twice the oMSD.

\begin{figure}[bh]
\includegraphics[width=0.73\linewidth]{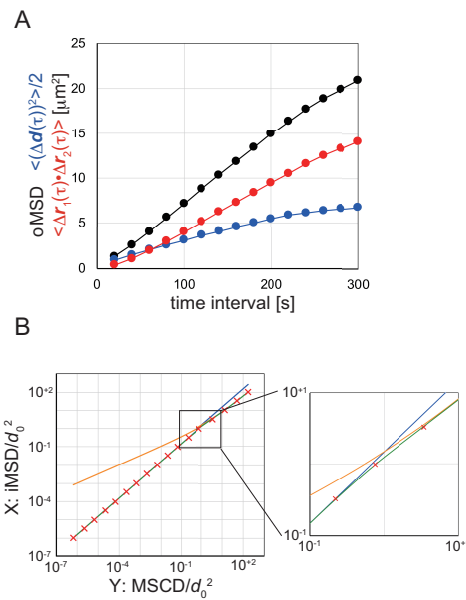}
\caption{Evaluation of chromatin motion independent of nucleus motion. (A) Evaluation of the contribution of nuclear movement. Data from the 8-cell stage were analyzed. For the other stages, see Sup. Fig. S2. oMSD ("observed MSD", black) is the MSD calculated from the simple tracking of the spot, and may include the effects of the nuclear movement. Blue is the term $\frac{\langle (\Delta {\vec d}(\tau))^2 \rangle}{2}$. Red is the correlation term $\langle \Delta {\vec r}_1(\tau) \cdot  \Delta {\vec r}_2(\tau) \rangle$, reflecting the contribution of nuclear movement. (B) Conversion formula from MSCD to iMSD ("intrinsic MSD"), excluding the effect of nuclear movement. The blue line is $X = f(Y) = \frac{3}{2}Y$. The orange line is $X = g(Y) = \frac{Y}{2}+\sqrt{Y}$. The green line is $X = [{f(Y)}^{-4}+{g(Y)}^{-4}]^{-\frac{1}{4}}$. The red cross indicates the results of the simulation.}
\label{Fig3}
\end{figure}

In Fig. 3, we compare the three terms in Eq.~(\ref{MSCDV}) measured in our experiment (Fig. 3A, Sup. Fig. S2). 
It is evident that the observed MSD is dominated by the correlation $\langle \Delta {\vec r}_1(\tau) \cdot  \Delta {\vec r}_2(\tau) \rangle$, which increases almost linearly with time. Note that the origin of measured correlation is not the coherent motion of the chromosomes inside the nucleus~\cite{Zidovska_PNAS2013,Michieletto_MacrLett2021}. Rather, the above observation, in particular the linear scaling of the correlation, indicates that the two tracked loci move together with the nucleus undergoing Brownian motion. 

{\it Mapping to intrinsic MSD.---}
Although our use of MSCD is motivated by experiments, its relationship with an "intrinsic MSD (iMSD)", the movement of the loci excluding the contribution of the nuclear movement, must be clarified. In addition to its familiarity, the latter has an advantage in that it is more compatible with theoretical analysis.
We obtained a formula relating iMSD and MSCD as follows: $iMSD = [{f(MSCD)}^{-4}+{g(MSCD)}^{-4}]^{-\frac{1}{4}}$, where $f(Y) = \frac{3}{2} Y$ and $g(Y) = \frac{Y}{2}+\sqrt{Y}$ (Fig. 3B; see SI for derivation and simulation).
Because MSCD is an index of chromatin motion independent of the movement of the nucleus, the iMSD obtained from the MSCD is also independent of the movement of the nucleus.

By applying this formula to the MSCD data obtained by two-point tracking, we obtained the iMSD of loci, which results entirely from the intrinsic chromatin dynamics (Fig. 4A).  Similar to the MSCD, we found that the MSD exponent $\alpha$ is relatively insensitive to nuclear size (Fig. 4B, blue). The average $\alpha$ was $0.44 \pm 0.02$ (95\% confidence interval [CI]). By fitting MSD from nuclei grouped according to their size to the formula ~(\ref{eq:MSD}) with $\alpha = 0.44$, we could obtain the mobility $A$ (i.e. MSD coefficient) as a function of nuclear size $R$ (Fig. 4B). This analysis yields the relation 
\begin{eqnarray}
A \sim R^{2.0 \pm 0.6} ,
\label{A_R}
\end{eqnarray}
which quantitatively describes the mobility reduction in smaller nuclei (Fig. 4B, red dotted line).

\begin{figure}[bh]
\includegraphics[width=0.73\linewidth]{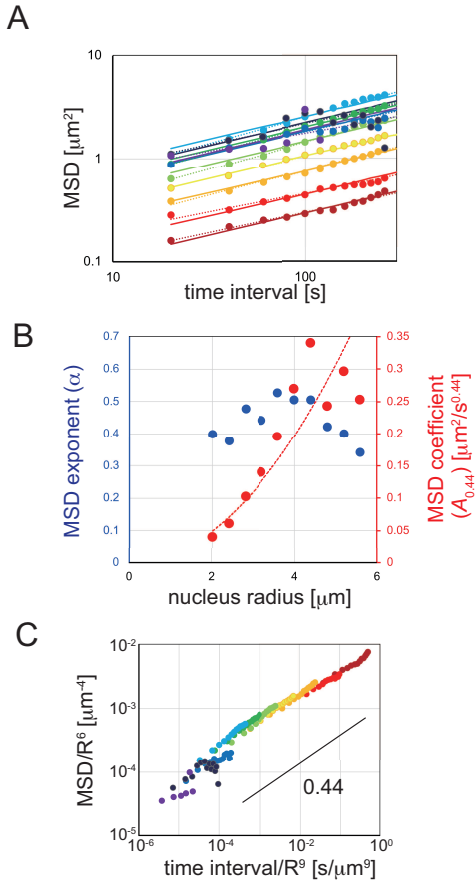}
\caption{Analysis of intrinsic MSD (iMSD) deduced from MSCD using the proposed conversion formula (see Fig. 3). (A) MSD of different nuclei sizes. The color is the same as in Fig. 1A, while we have an additional group with radius = 5.6 $\mu m$ (purple). The solid lines represent fitting to Eq. (1), whereas the dotted lines are fitting to Eq. (1) with $\alpha$ fixed to 0.44. (B) MSD exponent (blue) and MSD coefficient (red) of each group. No significant correlation was found between the MSD exponent ($\alpha$) and the nuclear radius ($p = 0.7$), whereas the correlation was significant between the MSD coefficient ($A_{0.44}$) and the nuclear radius ($p = 9 \times 10^{-4}$). Fitting of the MSD coefficient against the nuclear radius resulted in the formula $A = 0.0124 \times R^{1.99}$. (C) The master curve of MSD constructed by time-nuclear size superposition (double logarithmic scale). Rescaling of MSD (vertical axis) and time (horizontal axis) is performed according to the Lin-Noolandi conjecture with $\nu=1/2$. Rescaling with other models of entanglement are shown in Sup. Fig. S3. The color code is the same as in Fig. 4A.}
\label{Fig4}
\end{figure}

{\it Topological constraints are tighter in a smaller nucleus.---}
We now discuss the physics underlying the nuclear size dependence of chromatin mobility. During embryogenesis, the nuclear size becomes smaller almost in proportion to the cell size~\cite{Hara_CB2009}, while the amount of genetic material in individual nuclei remains constant.
At these stages, chromatin is distributed almost uniformly in the nucleus~\cite{Arai_SR2017}.
This means that the chromatin concentration becomes higher with a smaller nucleus. It might appear that the lower mobility at higher chromatin concentration is intuitively obvious because, by invoking the Stokes-Einstein law for a colloid particle in a viscous fluid, viscosity might become higher at higher chromatin concentrations.
Despite the intuitive nature of this understanding, there is a deficit in the above naive argument, in which one confuses the length scales associated with the viscosity. By convention, viscosity is a macroscopic concept; thus, in the present context, we have implicitly discussed viscosity at the nuclear scale (or the scale of the colloid particle), where one can describe the nucleoplasm as a continuum. However, the motion of chromatin loci themselves occurs in a more local environment. Invoking such a length-scale dependent local viscosity (or, more generally, viscoelastisity) often encountered in complex fluids, it is reasonable to assume that the chromatin loci move in a background fluid, whose viscosity $\eta$ is comparable to that of water.
In a simple polymer model such as the Rouse model, the mobility of monomers should not depend on the polymer concentration~\cite{deGennes_book, Rubinstein_book}.

To formulate the concentration-dependency of chromatin mobility, we focused on the TCs associated with the chain connectivity of the polymer. 
As chromatin chains are very long polymers, and are packed in the nucleus at high concentrations, there should be severe motion restrictions arising from the fact that these long polymers cannot freely pass through each other.
We emphasize here that the presence of TC does not necessarily indicates the chromatin in nucleus can be viewed as the entangled concentrated linear polymer solutions. Rather, the chromosome conformational capture experiments invoke the abundance of large loop structures on the order of Mbp~\cite{Liberman-Aiden_Sci2009}, which suggests the similarity with the dense solution of non-concatenated ring polymers~\cite{Rosa_PLCB2008, Halverson_RPP2014}. In the latter, rings are not entangled in a classical sense, but their conformation and dynamics are dictated by the immense TCs~\cite{Cates_JP1986, Sakaue_PRL2011, Smrek_JPCM2015, Rubinstein_Macro2016, Michieletto_PRL2017, Sakaue_SoftMatter2018, Landuzzi_PRR2020}. 

On purely dimensional grounds, one can write the MSDs of chromatin loci as
\begin{eqnarray}
{\rm MSD}(\tau) = a^2 \left(\frac{\tau}{\tau_e} \right)^{\alpha}
\label{MSD_1}
\end{eqnarray}
where $a$ is a characteristic length scale relevant to the process under consideration, $\tau_e$ is the corresponding time scale, and $\alpha$ is a nontrivial exponent characterizing the large-scale loci dynamics~\cite{Weber_PRL2010, Weber_PNAS2012, Bancaud_GR2013, Amitai_PRE2013, Vandebroek_PRE2015, Sakaue_SoftMatter2017, Tamm_PRL2018, Put_PRE2019}. Over long length and time scales, the conformation and dynamics of chromatin will be dominated by TC~\cite{Rosa_PLCB2008, Halverson_RPP2014, Cates_JP1986, Sakaue_PRL2011, Smrek_JPCM2015, Rubinstein_Macro2016, Sakaue_SoftMatter2018, Landuzzi_PRR2020, Michieletto_PRL2017}, which set our basic length scale to be a so-called tube diameter~\cite{deGennes_book, Rubinstein_book}. In a larger length scale ($r > a$), the chromatin will exhibit a non-trivial conformation, which also affects the motion of chromatin loci at long time scale ($t > \tau_e$).

Once we consider the TCs, the length scale $a$ depends on the concentration $c$ of chromatin monomers. Assuming that the short scale ($r<a$) conformation of chromatin obeys random walk statistics, one can then follow the standard argument in polymer physics to find 
$a = b N_e^{1/2} \sim c^{-1}$, where $N_e \sim c^{-2}$ is the corresponding genomic distance with $b$ being the size of chromatin monomers~\cite{Lin_Macro1987, Kavassalis_PRL1987, Uchida_JCP2008} (see the SI for more details).
Similarly, the concentration dependence of the time scale $\tau_e$ at the scale of the tube diameter is evaluated as 
$\tau_e \simeq (\eta \xi^3/k_BT) (N_e/g)^2 \sim c^{-3}$, where $k_BT$ is the thermal energy and $\xi \simeq b g^{1/2} \sim c^{-1}$ is the so-called mesh size, above which the solvent-mediated hydrodynamic interactions are screened~\cite{deGennes_book, Rubinstein_book}.
Substituting these relations into Eq.~(\ref{MSD_1}), we obtained the concentration dependence in MSD of chromatin loci
\begin{eqnarray}
{\rm MSD} \sim c^{3\alpha -2}. 
\label{MSD_2}
\end{eqnarray}
Because $c \sim R^{-3}$, this leads to ${\rm MSD} \sim R^{6-9\alpha}$. Using the experimentally obtained exponent $\alpha = 0.44$, our theory predicts that the chromatin mobility depends on the nuclear size as $A \sim R^{2.0}$, which agrees well with experimental observations (Fig. 4B).  Other models of entanglement predict qualitatively similar results, but with different exponents for $R$ dependence; see the SI for comparisons with other models of entanglement.

{\it Time-nuclear size superposition. ---}
Our basic formula~(\ref{MSD_1}) for MSD suggests that the distinct MSD vs. $\tau$ data from different nuclear sizes can be collapsed to a single master curve upon proper rescaling of the length and time. 
\begin{eqnarray}
\widetilde{{\rm MSD}} = {\tilde \tau}^{\alpha}
\label{MSD_master}
\end{eqnarray}
where $\widetilde{{\rm MSD}} \equiv {\rm MSD}/a^2 \sim {\rm MSD}/R^6$ and ${\tilde \tau} \equiv t/\tau_e \sim t/R^9$. In Fig. 4C and Sup. Fig. S3, we collect MSD data obtained from various nuclear sizes $R$, and construct the rescaled plot on a double logarithmic scale. The observed excellent collapse into a master curve (Fig. 4C and Sup. Fig. S3) supports the proposed physical picture. 
This analysis scheme provides an efficient way to extend the time range, which is usually limited by factors such as experimental apparatus and experimental conditions. This phenomenon is analogous to the so-called time-temperature superposition principle adopted in rheological measurement to obtain the viscoelastic modulus over a wide time (or frequency) range from measurements of a shorter time range at various temperatures~\cite{Larson_book}. While rescaling (i.e., a ``shift" in logarithmic scale) factors in rheological time-temperature superposition originate from the temperature dependence of the friction coefficient, they arise from, in our case, the concentration dependence of the topological constraints.  The experimentally accessible time range for tracking chromatin loci is limited to $20 \sim 300$ s in the present study, but the rescaled plot covers an effective time range spanning almost five orders of magnitude.

{\it Perspectives ---}
Recent studies have shown that nuclear size affects the formation of nucleoli~\cite{Weber_CB2015} and chromosome condensation~\cite{Hara_MBoC2013}. 
We have shown that nuclear size also affects the intrinsic mobility of chromatin in early embryos {\it C. elegans}.
We have succeeded in establishing a theory to relate chromatin movement and nuclear size, as ${\rm MSD} \sim R^{6-9\alpha}$, taking the TCs of polymers into account.
Since nuclear size reduces with cell stage, we speculate that the reduction in nuclear size has a regulatory role in controlling nuclear functions during embryogenesis. 
We expect that the present method will have versatile applicability to the quantification of intrinsic chromatin mobility in live cells, and it is an interesting challenge to clarify whether the proposed mechanism based on the TC is at work to control chromatin dynamics in other types of cells.

We thank Kazuko Ohishi for technical assistance, and Masaki Sasai, Tetsuya Yamamoto, Masaaki Sugiyama, and Hiroshi Kimura from the research group of "Chromatin Potential" area, Yuta Shimamoto and the members of the Kimura lab and Sakaue lab for discussion. This project was supported by JSPS KAKENHI (grant numbers: JP18H05529 to T.S. and A.K., and 20114008, JP16H05119, JP16H00816 to A.K.).

\end{document}


\title{SI: Supplementary Information for

Quantifying the mobility of chromatin during embryogenesis: Nuclear size matters}

\author{Aiya K. Yesbolatova}
\affiliation{Department of Genetics, School of Life Science, The Graduate University for Advanced Studies, SOKENDAI, Mishima 411-8540, Japan}
\affiliation{Cell Architecture Laboratory, Department of Chromosome Science, National Institute of Genetics, Mishima 411-8540, Japan}
\author{Ritsuko Arai}
\affiliation{Cell Architecture Laboratory, Department of Chromosome Science, National Institute of Genetics, Mishima 411-8540, Japan}
\affiliation{Present address: Department of Anatomy and Histology, Fukushima Medical University, School of Medicine, 1 Hikarigaoka, Fukushima, 960-1295, Japan}
\author{Takahiro Sakaue}
\email{sakaue@phys.aoyamna.ac.jp}
\affiliation{Department of Physics and Mathematics, Aoyama Gakuin University, 5-10-1 Fuchinobe, Chuo-ku, Sagamihara, Kanagawa 252-5258, Japan}
\author{Akatsuki Kimura}
\email{akkimura@nig.ac.jp}
\affiliation{Department of Genetics, School of Life Science, The Graduate University for Advanced Studies, SOKENDAI, Mishima 411-8540, Japan}
\affiliation{Cell Architecture Laboratory, Department of Chromosome Science, National Institute of Genetics, Mishima 411-8540, Japan}


\maketitle

{\bf METHODS}
\vspace{1cm}

{\it Live cell imaging and tracking of {\it lacO} spots.---}
Experimental data of the tracking of the {\it lacO} spots in live {\it Caenorhabiditis elegans} embryos were derived partly from our previous study~\cite{Arai_SR2017}, and from the results of the present study (Table ~\ref{nucleusnumber}). We used the AV221 strain, in which {\it lacO} repeats and GFP::LacI encoding DNA elements were integrated into the {\it C. elegans} genome~\cite{Bilgir_G32013} for new sets of experiments. We previously confirmed that AV221 and CAL0872 strains, in which the {\it lacO} repeats are integrated at an independent locus, show similar moving behavior during early embryogenesis~\cite{Arai_SR2017}. Therefore, we believe that the behavior we observe is not specific to the specific locus of the chromosome, but reflects a general trend. 

	We followed our previous procedures for the maintenance of the strains, image acquisition, and tracking, as described in~\cite{Arai_SR2017}. In brief, the strains were maintained using standard procedures~\cite{Brenner_G1974}. Embryos were isolated and observed using spinning-disk confocal microscope systems (Yokogawa, Tokyo, Japan) equipped with a UPlanSApo 100$\times$/1.40 NA objective (Olympus, Tokyo, Japan) and iXon 897 EMCCD camera (Oxford Instruments, Abingdon, UK). The fluorescent signals from GFP::LacI fustion protein bound to the { \\ it lacO} sites (“{\it lacO} spots”) were visualized by exciting the GFP with a 488 nm laser at 25 °C. Confocal sections were obtained with 1- or 2-$\mu m$ intervals for every 20 s. The positions of the {\it lacO} spots were tracked manually using the ImarisTrack software (Oxford Instruments). The radii of the nuclei were measured at the first and last frame of each tracking, and the averaged values were used as the radii of the nuclei.
	
	Knockdown of the genes {\it ima-3} and {\it C27D9.1} was conducted through the injection method of RNA-mediated interference (RNAi), as described previously~\cite{Hara_MBoC2013}.

\begin{table}
\caption{\bf Table S1: the number of nuclei tracked in this study (the numbers of 8-cell stage nuclei are parenthesized)}
\label{nucleusnumber}
\begin{center}
\begin{tabular}{llrrrrr} \toprule
label & range & total & Arai et al. 2017 &  & this study &  \\
  &   &   &   & control & \it{ima-3} & \it{C27D9.1} \\ 
2.0 & 1.8 - 2.2 & 54 (1) & 49 (0) & 4 (0) & 1 (1) & 0 (0) \\
2.4 & 2.2 - 2.6 & 23 (12) & 13 (5) & 3 (0) & 7 (7) & 0 (0) \\
2.8 & 2.6 - 3.0 & 92 (36) & 16 (6) & 36 (8) & 32 (18) & 8 (4) \\
3.2 & 3.0 - 3.4 & 138 (50) & 25 (16) & 60 (17) & 39 (9) & 14 (8) \\
3.6 & 3.4 - 3.8 & 146 (35) & 43 (14) & 57 (12) & 30 (1) & 16 (8) \\
4.0 & 3.8 - 4.2 & 102 (12) & 27 (1) & 50 (3) & 13 (0) & 12 (8) \\
4.4 & 4.2 - 4.6 & 50 (4) & 18 (0) & 25 (0) & 1 (0) & 6 (4) \\
4.8 & 4.6 - 5.0 & 16 (0) & 1 (0) & 14 (0) & 0 (0) & 1 (0) \\
5.2 & 5.0 - 5.4 & 8 (0) & 0 (0) & 8 (0) & 0 (0) & 0 (0) \\
5.6 & 5.4 - 5.8 & 3 (0) & 0 (0) & 3 (0) & 0 (0) & 0 (0) \\ 
total &  & 632 (150) & 192 (42) & 260 (40) & 123 (36) & 57 (32) \\ 
\end{tabular}
\end{center}
\end{table}

\vspace{1cm}
{\it Estimation of intrinsic MSD (iMSD) from the MSCD.---}
Here, we first provide an analytical expression to connect MSD and MSCD.
Let ${\vec d}_0 = {\vec r}_{2,0} - {\vec r}_{1,0}$ be an initial vector connecting the two loci. After arbitrary displacements $\Delta  {\vec r}_{i,0}$ of two loci ($i=1$ or $2$), this vector changes to ${\vec d} = {\vec d}_0 + \Delta{\vec r}_{2} - \Delta{\vec r}_{1}$. By introducing the quantity $P = |\Delta {\vec r}_2 - \Delta {\vec r}_1|/d_0$, the normalized change in distance is expressed as $\Delta d/d_0 = \sqrt{1 + P^2 + 2 P \cos{\theta}} -1$, where $\theta$ is the angle between ${\vec d}_0$ and $\Delta{\vec r}_{2} - \Delta{\vec r}_{1}$.
Assuming the isotropy and the negligible contribution of the displacement correlation $\langle \Delta {\vec r}_1 \cdot \Delta {\vec
r}_2 \rangle /MSD \ll 1$, $\langle P \cos{\theta} \rangle = \langle P \rangle \langle \cos{\theta} \rangle =0$,  we find a relation between the MSCD and the MSD
\begin{eqnarray}
Y &=& 2\left[ 1+  X -  \langle \sqrt{1 + P^2 +  2 P \cos{\theta}} \rangle \right] \nonumber \\
&\simeq &\left\{
\begin{array}{ll}
\frac{2}{3} X  & ( {\rm small} \ P )  \\
2\left[ 1 +X - \sqrt{1+ 2 X}  \right] & ( {\rm preaverage} ), 
\end{array}
\right.
\label{eq:MSCSD_2}
\end{eqnarray}
where $Y \equiv {\rm MSCD}/d_0^2$ and $X \equiv {\rm MSD}/d_0^2$ are the MSCD and the MSD normalized by $d_0^2$, respectively. The top or bottom in the final expressions are obtained by an expansion around $P=0$ (up to order ${\mathcal O}(P^4)$) or the preaveraging approximation $\langle \sqrt{1+P^2 + 2 P \cos{\theta}}\rangle \rightarrow \sqrt{1+\langle P^2\rangle + 2 \langle P \cos{\theta} \rangle}$. From Eq. ~(\ref{eq:MSCSD_2}), we obtain: 
\begin{eqnarray}
X &\simeq &\left\{
\begin{array}{ll}
f(Y) = \frac{3}{2} Y  & ( {\rm small} \ P ),  \\
g(Y) = \frac{Y}{2}+\sqrt{Y} & ( {\rm preaverage} )
\end{array}
\right.
\label{MSCSD_3}
\end{eqnarray}

	To confirm the analytical solutions and obtain an appropriate interpolation formula, we simulated the relationship between MSD and MSCD. In this simulation, we first placed two spots at $r_{1,0} = (+d_0/2, 0, 0)$ and $r_{2,0} = (-d_0/2, 0, 0)$, where $d_0$ is the initial distance between the two spots. Then, we moved the two spots toward an independent, random direction with a length of $l$. The resultant coordinates of the two spots are ${\vec r_1}$ and ${\vec r_2}$, respectively. In this setting, the squared displacement (SD) of the spots is $l^2$ and the squared change in distance (SCD) is $(|{\vec r_2} - {\vec r_1}| - d_0)^2$. In our experimental measurements, ${\rm SCD}/d_0^2$ ranged between $~10^{-6}$ and $~10^2$. Our analytical relationship between SD and SCD indicated that SCD and SD are of similar order, so we simulated settings where $l^2/d_0^2$ ranged between $10^{-6}$ and $10^2$. The single step, random movement of the two spots was simulated $10^5$ times for each combination of $l$ and $d_0$, and the averaged value of ${\rm SCD}/d_0^2 (=Y)$ was related to $l^2/d_0^2 (=X)$. The calculations were conducted using MATLAB software. The values obtained in the simulation agreed well with the analytically derived formula. Among the interpolation formula of $X = [f(Y)^{-p} + g(Y)^{-p}]^{-1/p} (p=1,2,3,4,5)$, the formula with $p=4$ gave the least sum of squared error with the simulation. Therefore, we adopted $X = [{f(Y)}^{-4}+{g(Y)}^{-4}]^{-\frac{1}{4}}$ as the formula to estimate MSD from MSCD (Fig. 3B). Comparison between the formula and the numerical simulation showed a deviation of less than 3 \%.

\vspace{1cm}
{\it Quantification of MSCDs and MSDs.---}
The calculation of the MSCD from the tracking data was conducted as described in~\cite{Arai_SR2017}. First, the distance between the two spots, $d(s, t)$, at time $t$ in sample $s$, was calculated from the coordinates of the pair of {\it lacO} spots in each nucleus. The MSCD for a given time interval, $\tau$, was calculated by averaging SCD, which is $[d(s, t+\tau ) - d(s, t)]^2$, for all possible values of $s$ and $t$ for nuclei whose size is in the range of interest. 

	To estimate intrinsic MSD (iMSD), the first individual SD was estimated as $SD = [f(SCD)^{-4} + g(SCD)^{-4}]^{-1/4}$ (see the previous section for $f(Y)$ and $g(Y)$). The iMSD for a given time interval, $\tau$, was calculated by averaging the SD for all possible values of $s$ and $t$ for nuclei whose size is in the range of interest. 

Calculations of observed MSD (oMSD), including the effect of nuclear movement, and $\frac{\langle (\Delta {\vec d}(\tau))^2 \rangle}{2}$ from the experiments in Fig. 3A was conducted according to the definition of oMSD and $\frac{\langle (\Delta {\vec d}(\tau))^2 \rangle}{2}$. The correlation term, $\langle \Delta {\vec r}_1(\tau) \cdot  \Delta {\vec r}_2(\tau) \rangle$, in Fig. 3A was calculated by subtracting $\frac{\langle (\Delta {\vec d}(\tau))^2 \rangle}{2}$ from the oMSD.
All the calculations were conducted using MATLAB software.

\vspace{1cm}
{\it Fitting MSCD/MSD to exponential functions, and statistical evaluation of the constants.---}
To fit the MSCD to $MSCD(\tau) = B\tau ^{\beta}$ (Eq. 3), or MSD to $MSD(\tau) = A\tau ^{\alpha}$ (Eq. 1), linear regression analysis was applied to pairs of $log(MSCD(\tau))$ or $log(MSD(\tau))$ and $log(\tau)$, and the slope and the intercept corresponded to $\beta$ or $\alpha$, and $exp(B)$ or $exp(A)$, respectively. For the larger $\tau$, fewer pairs exist to calculate SCD and SD, we weighted pairs of $log(MSCD(\tau))$ or $log(MSD(\tau))$ and $log(\tau)$ by the number of existing pairs. 

	To compare the mobility among different sized nuclei, we used the values of $B$ and $A$ as indices, because the values of $\beta$ and $\alpha$ take similar values among different classes of nuclear size. To compare the mobility with $B$ and $A$, we needed to use common values of $\beta$ and $\alpha$. We adopted the mean value of $\beta$ and $\alpha$ as the common $\beta$ and $\alpha$. We then calculated the best fit $B$ and $A$ under the common $\beta$ and $\alpha$. In Figs. 1A and 4A, dotted lines show the fitting results with distinct $\beta$ and $\alpha$, and the solid lines are those with common $\beta$ and $\alpha$, respectively.
	
	The statistical significance of the correlation between MSCD/MSD exponents or MSCD/MSD coefficient and nuclear size (Figs. 1B, 2B and 4B) was evaluated by calculating $p$-values for testing the hypothesis of no correlation against that of a nonzero correlation by using the corr function of MATLAB software. The 95\% confidence interval for the mean of $\alpha$ was calculated assuming an unknown variance. The 95\% interval of the exponent of the nucleus radius, $R$, in Fig. 4C is calculated using the linear regression analysis tool in Microsoft Excel by fitting $log(B)$ against $log(R)$.

\vspace{1cm}
{\it Concentration dependence of tube diameter and corresponding time scale.---}
In a coarse-grained description, chromatin in the nucleus may be modeled as a solution of flexible polymers. Let the size of the chromatin monomer be $b$, and the overall chromatin monomer concentration $c$. An important length scale in a polymer solution is a mesh size $\xi$, which is determined from the condition $\xi \sim b g^{\nu}$ and $g/\xi^3 \sim c$, where $g$ is the number of monomers constituting the mesh~\cite{deGennes_book, Rubinstein_book}. The first relation describes the conformational characteristics of the sub-chain inside the mesh via the Flory exponent $\nu$ (its value unspecified for the moment), and the second condition represents that meshes are space-filling. Then, we find $\xi \sim b c^{\nu/(1-3\nu)}$ and $g \sim c^{1/(1-3\nu)}$.

The topological constraints may be described by the entanglement concept in polymer physics, in which the notion of tube diameter $a$ and the associated number $N_e$ of monomers  plays a key role. Because the entanglement strand is a random walk of $N_e/g$ meshes, the tube diameter $a$ is represented as $a \sim \xi (N_e/g)^{1/2}$. The concentration dependence of $a$ and $N_e$ is obtained by the so-called Lin-Noolandi conjecture $(c/N_e) a^3 = n_s$, which states that there is a constant number $n_s (\sim 10)$ of entanglement strands per entanglement volume $\sim a^3$~\cite{Lin_Macro1987, Kavassalis_PRL1987}. From this, one finds
\begin{eqnarray}
a \simeq n_s \xi \sim c^{\nu/(1-3\nu)}, \\
\label{eq:a}
N_e \simeq n_s^2 g \sim c^{1/(1-3\nu)}
\label{eq:N_e}
\end{eqnarray}
The relaxation time $\tau_e$ of the entanglement strand is evaluated by noting that the solvent-mediated hydrodynamic interactions are screened at the mesh scale, that is, the Zimm-like dynamics at $r<\xi$ and the Rouse dynamics at $r > \xi$~\cite{deGennes_book, Rubinstein_book}:
\begin{eqnarray}
\tau_e \simeq \frac{\eta \xi^3}{k_BT} \left( \frac{N_e}{g}\right)^2 \sim c^{3\nu/(1-3\nu)}
\label{eq:tau_e}
\end{eqnarray}
where $\eta$ and $k_BT$ are the solvent viscosity and thermal energy, respectively.

Another conjecture on the entanglement entity, which may be appropriate for the concentrated $\Theta$ solution, is that of Colby-Rubinstein, in which the concentration dependence of the tube diameter is determined by the relation $cb^3 \times ca^3 = const$. Here, $cb^3$ is the probability of binary contacts per monomer, and $ca^3$ is the number of monomers in the entanglement cube. Hence, the assumption behind the conjecture is the constant number of binary contacts per entanglement volume $a^3$. This leads to 
\begin{eqnarray}
a \sim c^{-2/3}, \\
\label{eq:a-2}
N_e  \sim c^{-4/3}
\label{eq:N_e-2}
\end{eqnarray}
owing to the ideal chain statistics in a concentrated $\Theta$ solution. Then, with the mesh size scaling with $\nu=1/2$ ($\xi \sim c^{-1}$, $g \sim c^{-2}$), Eq.~\ref{eq:tau_e} gives 
\begin{eqnarray}
\tau_e \sim c^{-5/3} . 
\label{eq:tau_e-2}
\end{eqnarray}

\vspace{1cm}
{\it Concentration dependence of MSD and its master curve---}
The concentration dependence of MSD depends on the model of entanglement through the concentration dependence of $a$ and $\tau_e$. 
By employing the Lin-Noolandi conjecture, we substitute Eqs.~(\ref{eq:a}) and~(\ref{eq:tau_e}) in the MSD formula 
\begin{eqnarray}
{\rm MSD} = a^2 (t/\tau_e)^{\alpha}
\label{MSD_formula}
\end{eqnarray}
(see Eq. (8) in the main text), we obtain
\begin{eqnarray}
{\rm MSD} = b^2 n_s^{2-2\alpha}\left( \frac{\xi}{b}\right)^{2-3\alpha} \left( \frac{t}{\tau_b}\right)^{\alpha} \sim c^{-\frac{\nu(2-3\alpha)}{3\nu-1}}, 
\label{eq:MSD-1}
\end{eqnarray}
where $\tau_b = \eta b^3/k_BT$ is the monomer relaxation time scale.
In the concentrated solution, the ideal chain statistics are applied to all length scales. Thus, by setting $\nu=1/2$, we find ${\rm MSD} \sim c^{-(2-3\alpha)}$ (see Eq (9) in the main text). In the so-called semidilute solution regime, the excluded-volume effect is significant inside the mesh $\nu = 0.589 \cdots$. If we adopt Flory's mean-field value $\nu = 3/5$, we obtain ${\rm MSD} \sim c^{-3(2-3\alpha)/4}$.

If we employ the Colby-Rubinstein conjecture, combining Eqs~(\ref{eq:a-2}) and~(\ref{eq:tau_e-2}) with our MSD formula, we find its concentration dependence
\begin{eqnarray}
{\rm MSD} = b^2  \left( \frac{a}{b}\right)^{2-4\alpha} \left( \frac{\xi}{b}\right)^{\alpha} \left( \frac{t}{\tau_b}\right)^{\alpha} \sim c^{-\frac{4-5\alpha}{3}}, 
\label{eq:MSD-2}
\end{eqnarray}

The concentration $c$ dependence can be transformed into the form of the nuclear size $R$ dependence through $c \sim R^{-3}$. In Table~\ref{table:gamma}, we summarize the above results by listing the exponent $\gamma$, which is defined as ${\rm MSD} \sim R^{\gamma}$.

\begin{table}[h]
 \caption{Exponent $\gamma$ predicted by various entanglement models: Lin-Noolandi (L-N) conjecture with $\nu=1/2$ (concentration solution), $\nu=3/5$ (semi-dilute solution) and Colby-Rubinstein (C-R) conjecture.  In the bottom row, the experimentally measured MSD exponent $\alpha = 0.44$ is inserted. }
 \label{table:gamma}
 \centering
 \begin{tabular}{|c|c|c|c|}
 \hline
 & L-N & L-N   & C-R \\
 & ($\nu=1/2$) \  &($\nu=3/5$) \   & \\
 \hline \hline
 $\gamma$ & $6-9\alpha$  & $\frac{9}{4}(2-3\alpha)$ &  $4-5\alpha$ \\
 \hline
 $\gamma_{\alpha=0.44}$ & 2.04 & 1.53 & 1.8 \\ 
 \hline
 \end{tabular}
 \end{table}
 
As seen here, all the models agree at a qualitative level in the sense that they successfully captured the experimental observation that  the mobility algebraically increases with the nuclear size. Amongst them, the prediction $\gamma_{\alpha=0.44} = 2.04$ of the Lin-Noolandi conjecture with $\nu=1/2$ agrees with the experimental result $\gamma_{\alpha=0.44} = 1.99$ very well. 

To gain further insights into the nuclear size dependence of MSD, we now attempt to rescale the MSD data obtained from different nuclear sizes to construct a master curve. Our MSD formula Eq.~(\ref{MSD_formula}) suggests that all the data collapse into a single curve
\begin{eqnarray}
\widetilde{{\rm MSD}} = \tilde{\tau}^{\alpha}
\end{eqnarray}
where $\widetilde{{\rm MSD}} = {\rm MSD}/a^2$ and $ \tilde{\tau} = \tau/\tau_e$ are the rescaled MSD and time interval, respectively.
Nuclear size-dependent rescaling factors can be determined from the concentration dependence of $a$ (Eqs.~(\ref{eq:a}),~(\ref{eq:a-2})), and $\tau_e$ (Eqs.~(\ref{eq:tau_e}) and ~(\ref{eq:tau_e-2})). In Sup. Fig. S3, we show the resultant master curves, which are constructed according to the guidance of the three cases discussed. Evidently, the quality of collapse is best for the Lin-Noolandi conjecture with $\nu=1/2$, suggesting that it is the most appropriate model to quantify the concentration-dependent topological constraint for chromatin in the nucleus.

\newpage
{\bf SUPPLEMENTAL RESULTS }

\begin{figure}[bh]
\includegraphics[width=0.73\linewidth]{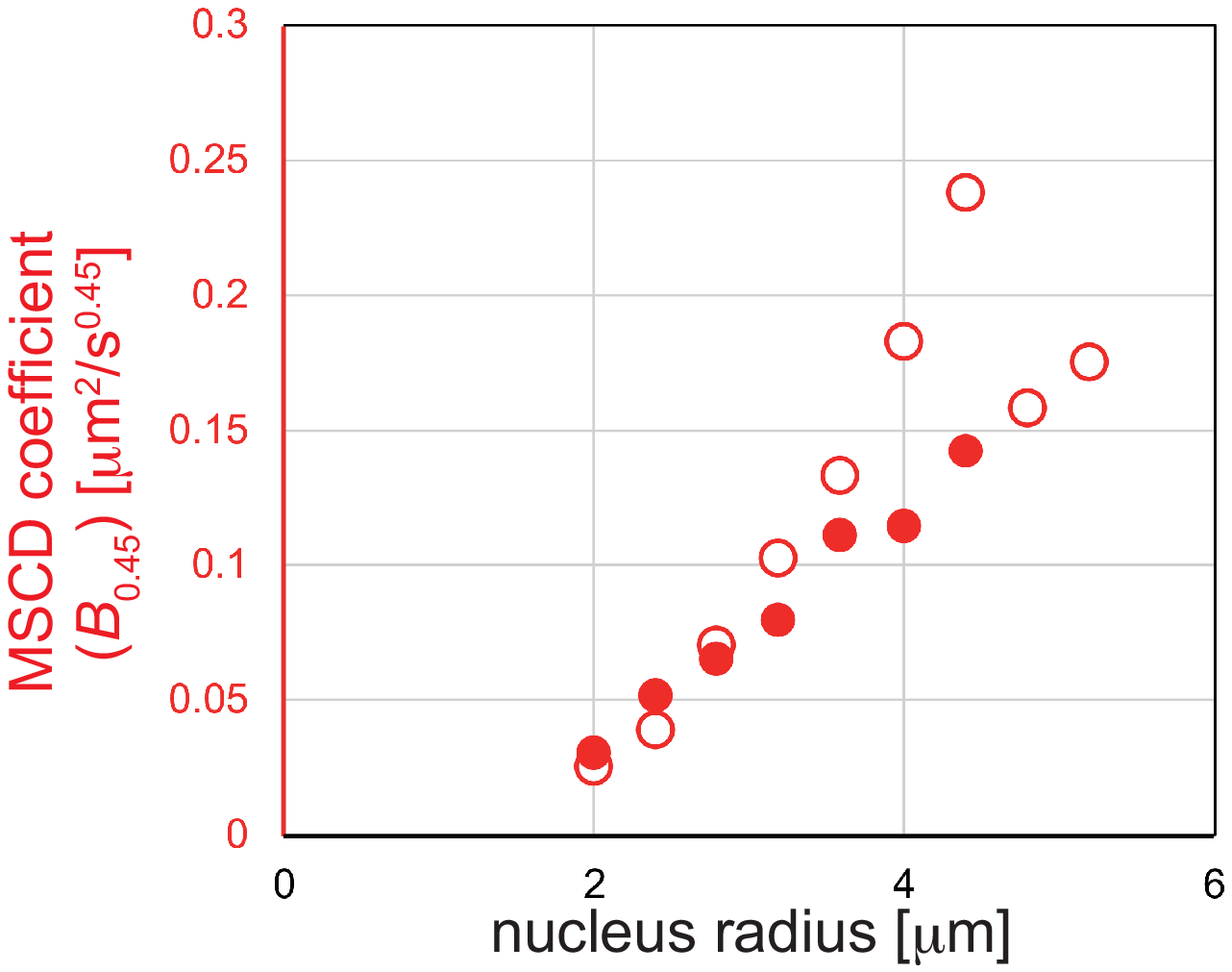}
\caption{FIG. S1: Comparison between the nuclear size dependency of the MSCD coefficient from the eight-cell stage (filled circle) and from all stages (blank circle).}
\label{FigS1}
\end{figure}

\begin{figure}[bh]
\includegraphics[width=0.73\linewidth]{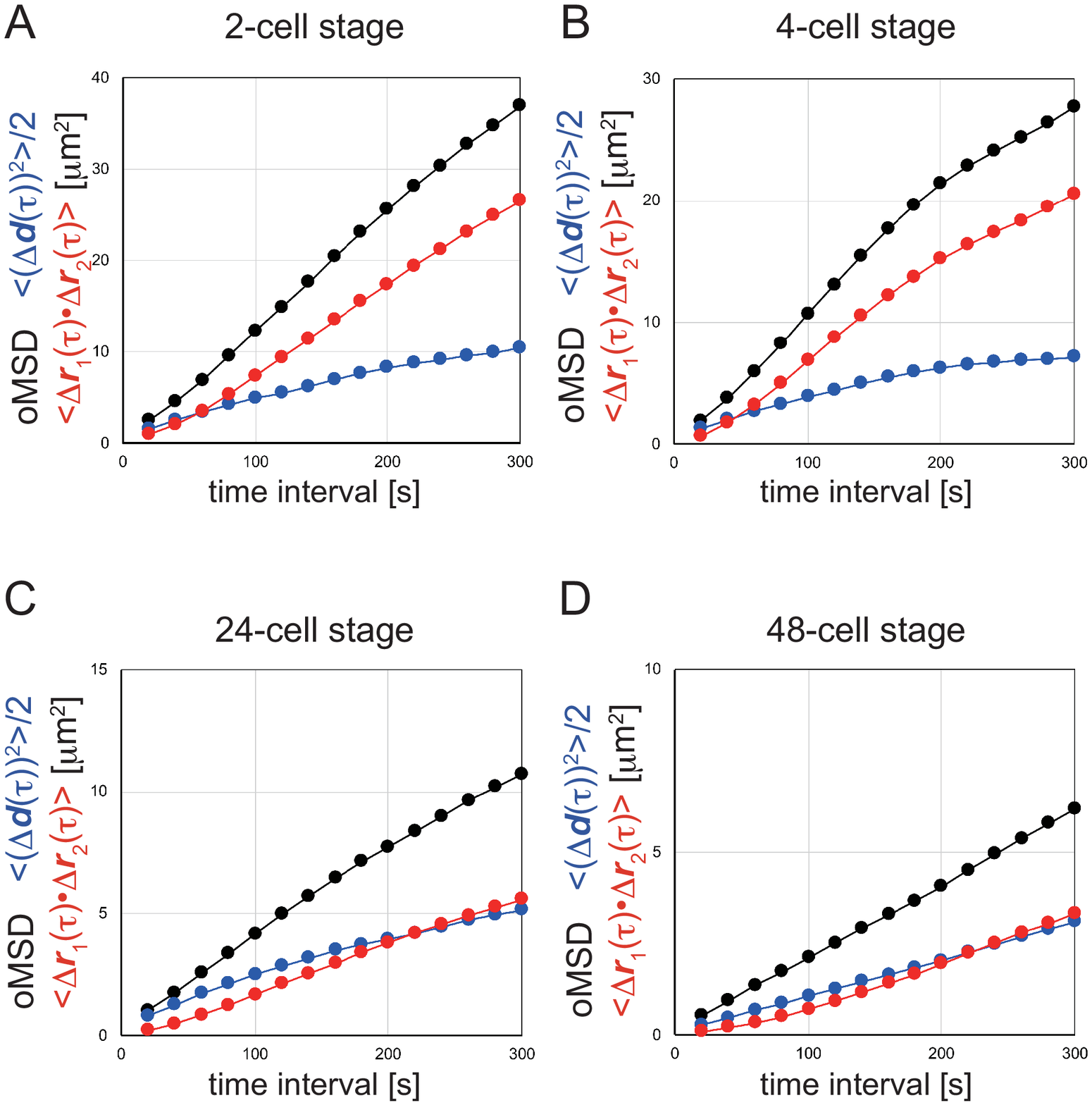}
\caption{FIG. S2: Contribution of nuclear movement in stages other than the eight-cell stage (Fig. 3A). (A) Two-cell stage. (B) Four-cell stage, (C) 24-cell stage, and (D) 48-cell stage. The color code is the same as in Fig. 3A.}
\label{FigS2}
\end{figure}

\begin{figure}[bh]
\includegraphics[width=0.73\linewidth]{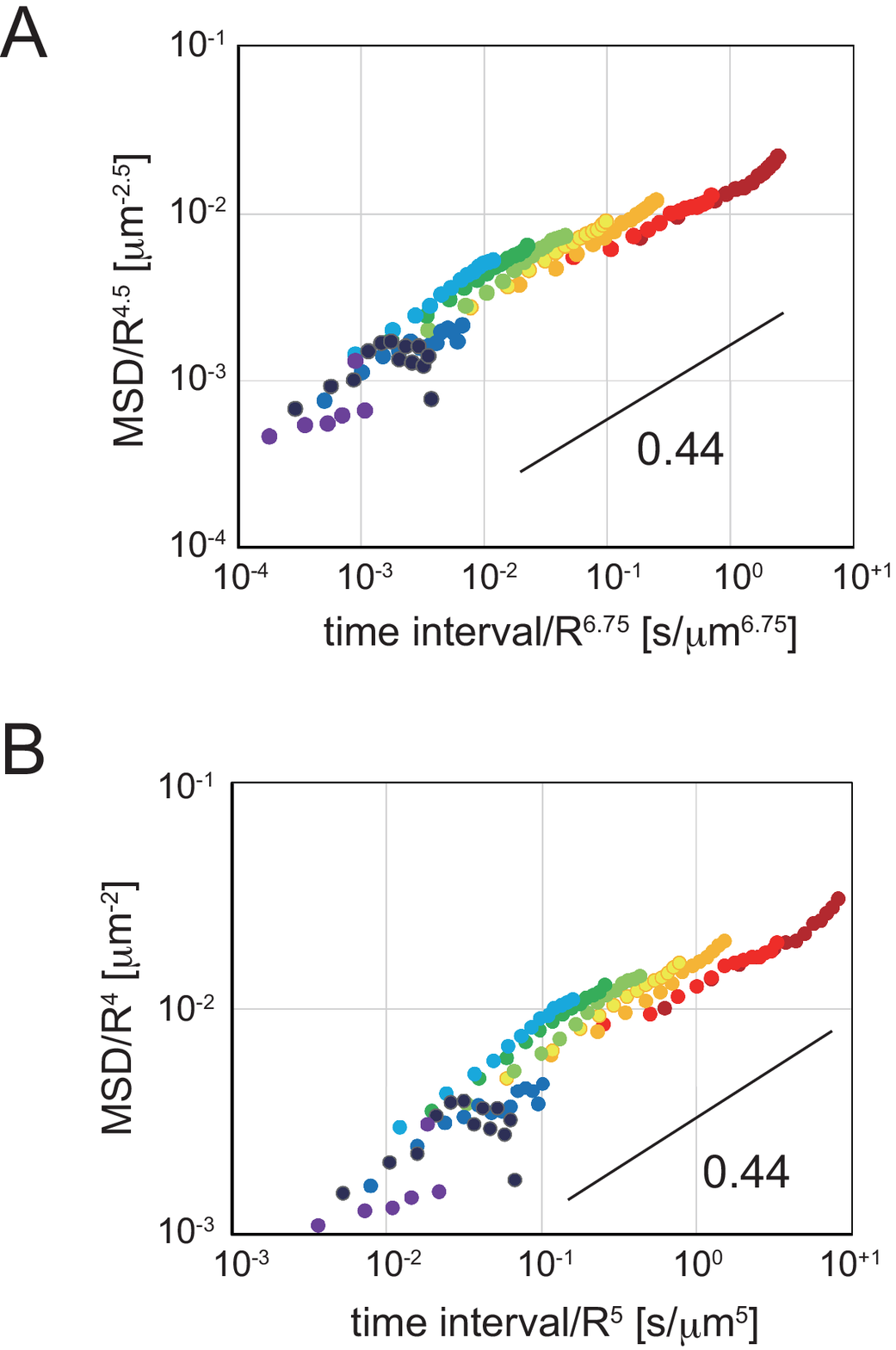}
\caption{FIG. S3: The master curve of MSD constructed by time-nuclear size superposition (double logarithmic scale), with rescaling other than Lin-Noolandi conjecture with $\nu=1/2$ shown in Fig. 4C. (A) Lin-Noolandi conjecture with $\nu=3/5$. (B) Colby-Rubinstein conjecture. The color code is the same as in Fig. 4C.}
\label{FigS3}
\end{figure}